\documentclass[journal]{IEEEtran}

\usepackage{lineno} % 这个包就是latex自带的添加行号的包了
\usepackage{amsmath,amsfonts}
\usepackage{array}
\usepackage{subcaption}
\usepackage{textcomp}
\usepackage{stfloats}
\usepackage{url}
\usepackage{verbatim}
\usepackage{graphicx}
\usepackage{cite}
\usepackage[bookmarksnumbered=true]{hyperref}
\hypersetup{
hidelinks,
colorlinks=true,
linkcolor=blue,
citecolor=blue,
urlcolor=black
}
\usepackage[linesnumbered,ruled,vlined]{algorithm2e}
\usepackage{bbm} % 字体bbl,空心字母，指示复数空间

\usepackage{amsthm} % theorem,lemma,proof环境

\usepackage{hyperref}
\usepackage{xcolor}
\hypersetup{
    colorlinks=true,      % 启用颜色超链接
    urlcolor=blue        % 设置URL超链接颜色为蓝色
}

\usepackage{enumerate} % 有序编号

\begin{document}

\newcounter{probcounter}
\newcommand{\refporbcounter}[1]{\refstepcounter{probcounter}\theprobcounter \label{#1}} % 新建命令，在正文中插入可引用的计数器，但是不能直接在公式环境中用

\title{LLM-Empowered Agentic AI for QoE-Aware Network Slicing Management in Industrial IoT}

\author{
Xudong Wang, 
Lei Feng,~\IEEEmembership{Member,~IEEE,}
Ruichen Zhang,
Fanqin Zhou,
Hongyang Du, \\
Wenjing Li,
Dusit Niyato,~\IEEEmembership{Fellow,~IEEE,}
Abbas Jamalipour,~\IEEEmembership{Fellow,~IEEE,}
Ping Zhang,~\IEEEmembership{Fellow,~IEEE}

\thanks{\textit{Corresponding author: Lei Feng}}
\thanks{
X. Wang, L. Feng, F. Zhou, W. Li, and P. Zhang are with the State Key Laboratory of Networking and Switching Technology, Beijing University of Posts and Telecommunications, Beijing, China, 100876 (e-mail: xdwang@bupt.edu.cn, fenglei@bupt.edu.cn, fqzhou2012@bupt.edu.cn, wjli@bupt.edu.cn, pzhang@bupt.edu.cn). }
\thanks{R. Zhang, and D. Niyato are with the College of Computing and Data Science,
Nanyang Technological University, Singapore (e-mail: ruichen.zhang@ntu.edu.sg, dniyato@ntu.edu.sg). }
\thanks{H. Du is with the Department of Electrical and Electronic Engineering, University of Hong Kong, Pok Fu Lam, Hong Kong SAR, China (e-mail: duhy@eee.hku.hk).}
\thanks{A. Jamalipour is with the School of Electrical and Computer Engineering,
University of Sydney, Australia (e-mail: a.jamalipour@ieee.org).}
}

% The paper headers
% \markboth{Journal of \LaTeX\ Class Files,~Vol.~14, No.~8, August~2021}%
% {Shell \MakeLowercase{\textit{et al.}}: A Sample Article Using IEEEtran.cls for IEEE Journals}

%\IEEEpubid{0000--0000/00\$00.00~\copyright~2021 IEEE}
% Remember, if you use this you must call \IEEEpubidadjcol in the second
% column for its text to clear the IEEEpubid mark.

\maketitle

\begin{abstract}

The Industrial Internet of Things (IIoT) requires networks that deliver ultra-low latency, high reliability, and cost efficiency, which traditional optimization methods and deep reinforcement learning (DRL)-based approaches struggle to provide under dynamic and heterogeneous workloads. To address this gap, large language model (LLM)-empowered agentic AI has emerged as a promising paradigm, integrating reasoning, planning, and adaptation to enable QoE-aware network management. In this paper, we explore the integration of agentic AI into QoE-aware network slicing for IIoT.
We first review the network slicing management architecture, QoE metrics for IIoT applications, and the challenges of dynamically managing heterogeneous network slices, while highlighting the motivations and advantages of adopting agentic AI. We then present the workflow of agentic AI-based slicing management, illustrating the full lifecycle of AI agents from processing slice requests to constructing slice instances and performing dynamic adjustments. Furthermore, we propose an LLM-empowered agentic AI approach for slicing management, which integrates a retrieval-augmented generation (RAG) module for semantic intent inference, a DRL-based orchestrator for slicing configuration, and an incremental memory mechanism for continual learning and adaptation.
Through a case study on heterogeneous slice management, we demonstrate that the proposed approach significantly outperforms other baselines in balancing latency, reliability, and cost, and achieves up to a 19\% improvement in slice availability ratio.
\end{abstract}

% \begin{IEEEkeywords}
% Generative Artificial Intelligence, Wireless Hallucination, Diffusion Model, Channel Estimation
% \end{IEEEkeywords}

\section{Introduction}

The Industrial Internet of Things (IIoT) has emerged as a cornerstone of modern industrial systems, enabling seamless integration of sensing, communication, and intelligent control across large-scale production environments~\cite{10852212}. Such systems impose stringent requirements on the underlying communication infrastructure, including ultra-low latency, high reliability, and cost efficiency, to support mission-critical operations and ensure industrial-grade quality of service. Network slicing (NS) has been widely recognized as a key enabling technology to meet these requirements by creating multiple virtualized and isolated logical networks over a shared physical infrastructure~\cite{9846944}. Each slice can be tailored to serve a specific use case with differentiated service-level guarantees. However, despite its promising potential, the dynamic management of network slicing in IIoT remains highly challenging, particularly when slices with heterogeneous quality of experience (QoE) demands arrive concurrently. These challenges highlight the urgent need for advanced intelligence in orchestrating QoE-aware slice allocation and adaptation.

To address these challenges, various approaches have been investigated in recent years. Conventional heuristic-based and optimization-driven methods have been widely applied to network slicing management in IIoT, offering tractable solutions for static or semi-dynamic scenarios. More recently, decision-making artificial intelligence techniques, particularly deep reinforcement learning (DRL), have been employed to automate slice orchestration by learning optimal policies from interaction with dynamic environments~\cite{9846944}. For example, the authors in~\cite{10561624} proposed a hierarchical DRL-based radio access network slicing scheme that leverages centralized training and distributed execution to enhance scalability while ensuring QoE in heterogeneous 6G environments. While these methods demonstrate improvements in adaptability, they often suffer from limited generalization, slow convergence, and brittleness when faced with highly dynamic and heterogeneous IIoT traffic patterns~\cite{11004012}. Consequently, neither traditional optimization nor decision-making AI is capable of meeting the requirements of real-time and QoE-aware slice management, motivating the exploration of more adaptive intelligence paradigms.

To overcome these limitations, large language model (LLM)-empowered agentic AI has recently attracted growing attention from both the AI and communications research communities~\cite{10638533}. Agentic AI, empowered by LLMs, refers to autonomous systems that integrate reasoning, planning, and acting capabilities to adaptively solve dynamic tasks in complex environments. Unlike conventional DRL agents, LLM-based agents exhibit enhanced contextual reasoning, self-reflection, and the ability to leverage external tools, which make them highly suitable for managing complex and uncertain network conditions~\cite{10638533}. Early studies have begun to explore the application of agentic AI in communication networks, demonstrating its potential in domains such as resource orchestration, anomaly detection, and intelligent automation. For instance, the authors in~\cite{jiang2025agentic} proposed a retrieval-augmented generation (RAG) transformer framework that harnesses LLM-driven agentic AI to enhance multi-UAV trajectory optimization in low-altitude networks. Nevertheless, the use of agentic AI for QoE-aware network slicing management in IIoT remains largely underexplored. This motivates our study, which investigates LLM-empowered agentic AI as a novel paradigm to enable adaptive, intelligent, and QoE-centric slice management in future industrial networks. The main contributions are summarized as follows:

\begin{itemize}
    \item We review the IIoT slicing management architecture and QoE metrics, analyze the unique challenges of QoE-aware network slicing management, and highlight the motivations and advantages of adopting agentic AI.
    \item We design a complete workflow for agentic AI-based slicing management. Then, we propose an LLM-empowered agentic AI approach for dynamic slicing management that integrates an RAG module for semantic intent inference, a DRL-based slice orchestrator for adaptive resource allocation, and an incremental memory mechanism for continual learning and self-refinement.
    \item We present a case study on heterogeneous slice management in IIoT, demonstrating that our proposed agentic AI approach significantly outperforms existing baselines by reducing latency and cost, while achieving up to a $19\%$ improvement in slice availability ratio.
\end{itemize}

\section{Rethinking Network Slicing in IIoT}

\subsection{Network Slicing Architecture for IIoT Systems}

\begin{figure*}
    \centering
    \includegraphics[width=0.8\linewidth]{figures/NS_architecture_0909.pdf}
    \caption{\small{The network slicing architecture for IIoT systems. End devices connect through 5G gNBs to a virtualized core network with VNF instances, OpenStack, and X86 infrastructure, orchestrated by an NS manager. The application layer supports AGVs, automatic control, VR/AR, and sensing via a project-oriented platform. The NS manager enables QoE management, resource allocation, and slice orchestration, while mapping KPIs into QoE metrics such as latency, reliability, economics, and throughput with representative IIoT applications.}}
    \label{NS_architecture}
\end{figure*}

% The deployment of network slicing in IIoT enables multiple logical networks on a shared physical infrastructure, each tailored to the stringent and diverse requirements of industrial applications~\cite{10551400}. IIoT environments feature highly heterogeneous service demands, including ultra-reliable low-latency communication (URLLC), massive machine-type communication (mMTC), and enhanced mobile broadband (eMBB). Fig.~\ref{NS_architecture} depicts a representative IIoT-oriented slicing architecture, emphasizing the interplay among terminals, radio access, core virtualization, slice management, and application platforms.
Typical IIoT services include ultra-reliable low-latency communication (URLLC), massive machine-type communication (mMTC), and enhanced mobile broadband (eMBB). Fig.~\ref{NS_architecture} illustrates an IIoT-oriented slicing architecture spanning terminals, access, core virtualization, slice management, and application platforms.

% \subsubsection{\textbf{End Devices and Service Differentiation}} At the edge of the system, diverse industrial terminals generate differentiated service demands. For instance, power automation terminals require URLLC to ensure safe and precise control of grid operations, while video monitoring terminals demand eMBB slices to support high-resolution, real-time video streams. These heterogeneous requirements drive the necessity for multi-slice instantiation within the same IIoT environment.

\subsubsection{\textbf{End Devices and Service Differentiation}} Industrial terminals generate heterogeneous traffic demands. For example, power automation devices rely on URLLC for safe grid control, while video surveillance terminals require eMBB for real-time high-definition streaming, motivating multi-slice coexistence within the same IIoT network.

\subsubsection{\textbf{Access and Transport Layer}} The terminals access the network via gNBs that provide slice-aware scheduling and traffic steering, ensuring that service flows are mapped to appropriate slices while preserving isolation and QoE guarantees.

\subsubsection{\textbf{Core Network Virtualization}} The core network (CN) is virtualized on X86 platforms using OpenStack, where slice-specific VNFs enable flexible, scalable, and rapid slice instantiation to accommodate dynamic IIoT demands.

\subsubsection{\textbf{Network Slicing Management}} A centralized NS management system orchestrates slice lifecycles and resource allocation, aligning physical infrastructure with logical slice templates to meet QoE requirements efficiently.

\subsubsection{\textbf{Application and Service Platform}} On top of the communication infrastructure, a project-oriented application enabling platform and embedded NS software stacks bridge industrial applications with the network, enabling QoE-aware operation and improved performance for industrial use cases.

% This architecture demonstrates how network slicing transforms a shared infrastructure into a set of service-specific virtual networks. Through slice orchestration and dynamic resource allocation, IIoT systems can flexibly provision differentiated network slices tailored to the latency, reliability, and throughput requirements of industrial processes such as real-time control, predictive maintenance, and digital twin systems.
Overall, network slicing transforms shared infrastructure into service-specific virtual networks, enabling flexible provisioning of latency, reliability, and throughput-aware slices for IIoT applications such as real-time control, predictive maintenance, and digital twins.

\subsection{QoE Metrics for IIoT NSs}

% The QoE in IIoT network slicing extends beyond conventional network-level Key Performance Indicators (KPIs). Instead of solely focusing on throughput or delay, QoE captures how well the deployed slices support industrial processes, user satisfaction, and economic efficiency. In IIoT systems, where both mission-critical control loops and large-scale monitoring tasks coexist, three major QoE metrics include latency, reliability, and economics, which is shown in Fig.~\ref{NS_architecture}.
In IIoT network slicing, the QoE extends beyond traditional network KPIs by reflecting how effectively slices support industrial processes, user satisfaction, and cost efficiency. Given the coexistence of mission-critical control and large-scale monitoring services, QoE in IIoT systems is primarily characterized by latency, reliability, and economics, as illustrated in Fig.~\ref{NS_architecture}.

% \subsubsection{\textbf{Latency}} Latency is one of the most critical QoE metrics in IIoT NSs, particularly for real-time control applications such as robotic motion control, automated guided vehicles, and smart grid protection~\cite{10620407}. URLLC slices must guarantee millisecond-level or even sub-millisecond response times to ensure safe and stable industrial operation. Any excessive delay directly degrades task completion time and threatens process continuity. Thus, latency is not only a KPI but also a direct reflection of perceived QoE in industrial automation.
\subsubsection{\textbf{Latency}} Latency is a dominant QoE metric for real-time industrial control applications, such as robotic automation and smart grid protection~\cite{10620407}. URLLC slices must ensure millisecond or sub-millisecond response times, as excessive delay directly threatens operational safety and process continuity. Consequently, latency represents not only a performance indicator but also a direct measure of perceived industrial QoE.

% \subsubsection{\textbf{Reliability}} In IIoT systems, reliability is closely tied to the degree of slice isolation. Stronger isolation among slices ensures that disturbances in one service (e.g., video monitoring traffic bursts in an eMBB slice) do not compromise the stability of mission-critical URLLC slices. Hence, the reliability perceived by industrial stakeholders depends on how effectively the slicing mechanism enforces logical and resource separation at the radio, core, and edge layers. From a QoE perspective, isolation-driven reliability guarantees deterministic service continuity, which is indispensable in safety-critical industrial processes~\cite{10475170}.

\subsubsection{\textbf{Reliability}} Reliability is closely linked to slice isolation in IIoT environments. Effective isolation prevents traffic fluctuations in non-critical slices from degrading URLLC services, thereby ensuring deterministic and continuous operation across radio, core, and edge layers. From a QoE perspective, such isolation-driven reliability is essential for safety-critical industrial processes~\cite{10475170}.

% \subsubsection{\textbf{Economics}} Beyond technical performance, economics is an emerging QoE dimension in IIoT slicing. Industrial operators increasingly focus on cost-efficiency, covering capital expenditure (CAPEX) for infrastructure and operational expenditure (OPEX) for orchestration and maintenance~\cite{10551400}. For instance, deploying redundant slices may enhance reliability but also increase costs. Thus, the economic aspect of QoE entails balancing service quality with deployment and management expenses to ensure slices remain technically viable and financially sustainable for large-scale IIoT adoption.

\subsubsection{\textbf{Economics}} Economics has emerged as a key QoE dimension, as industrial operators increasingly prioritize cost efficiency in both capital expenditure (CAPEX) and operational expenditure (OPEX)~\cite{10551400}. While techniques such as slice redundancy can enhance reliability, they also incur higher costs. Therefore, economic QoE reflects the tradeoff between service quality and deployment sustainability.

A comprehensive QoE assessment in IIoT requires integrating latency, reliability, and economics, with intelligent models to map network KPIs into perceived industrial outcomes for responsive, safe, and sustainable operations.

% \begin{figure*}
%     \centering
%     \includegraphics[width=0.8\linewidth]{figures/Figure1_20250226_GenAI_model.pdf}
%     \caption{\small{A summary of different types of GenAI models, potential causes and manifestations of hallucination.}}
%     \label{Figure1_GenAI_model_Hallucination}
% \end{figure*}

\subsection{Challenges of Dynamic NS Management}

The dynamic nature of IIoT imposes stringent demands on NS management, as static resource allocation is insufficient to cope with rapidly varying workloads and mission-critical service requirements. Effective dynamic slice management must address a series of technical and operational challenges that span multiple layers of the IIoT ecosystem.

\begin{itemize}
    \item \textbf{Modeling Task-Aware Joint Objectives:} Heterogeneous network slicing requires the QoE integration of different service-level agreements (SLA) performance into a unified optimization objective. However, IIoT tasks prioritize different factors such as latency, reliability, or economics, making static formulations ineffective. Dynamically assigning task-specific weights for joint performance remains a fundamental challenge.
    
    \item \textbf{Differentiating Fine-Grained Intents:} IIoT applications express diverse and evolving task semantics that influence slice behavior. Traditional intent classification lacks the granularity and flexibility needed to capture subtle distinctions, especially under ambiguous or natural-language task descriptions. Enabling fine-grained, interpretable intent recognition is essential for responsive and accurate slice management.
    
    \item \textbf{Adapting Slice Deployment in Dynamic Environments:} Maintaining QoE with expanding NS requirements demands timely adaptation of slice resources, including VNF placement and migration. The complexity of dynamic topologies, diverse latency constraints, and limited compute resources hinders real-time decision-making, especially under varying workloads.
\end{itemize}

Conventional DRL methods face intrinsic difficulties in addressing the above challenges.
These methods rely on predefined reward functions with static or heuristically tuned weights, which cannot effectively capture dynamically evolving and task-dependent QoE priorities in heterogeneous IIoT environments. Moreover, by operating solely on structured numerical states, conventional DRL lacks the capability to interpret high-level service intents and abstract operational goals, resulting in a semantic gap between industrial application requirements and low-level resource orchestration. To ensure QoE-aware, cost-efficient, and scalable orchestration, more intelligent, adaptive, and context-aware frameworks are needed. This motivates the use of agentic AI empowered by LLMs, which can reason over context, predict dynamics, and autonomously coordinate resources.

\subsection{Motivations of Using Agentic AI for IIoT Dynamic NSs}

Agentic AI technology represents a new paradigm where autonomous software agents possess goal-driven autonomy, reasoning, memory, coordination, and action execution capabilities~\cite{jiang2025large}. When combined with LLMs, these agents gain the additional ability to understand, interpret, and reason over semantic information from heterogeneous sources, such as network telemetry, industrial application intents, and operational manuals~\cite{jiang2025comprehensive}. In IIoT network slicing, this means agents can go beyond KPI-based optimization and instead manage slices with an explicit QoE-aware perspective. 

\begin{figure*}
    \centering
    \includegraphics[width=0.95\linewidth]{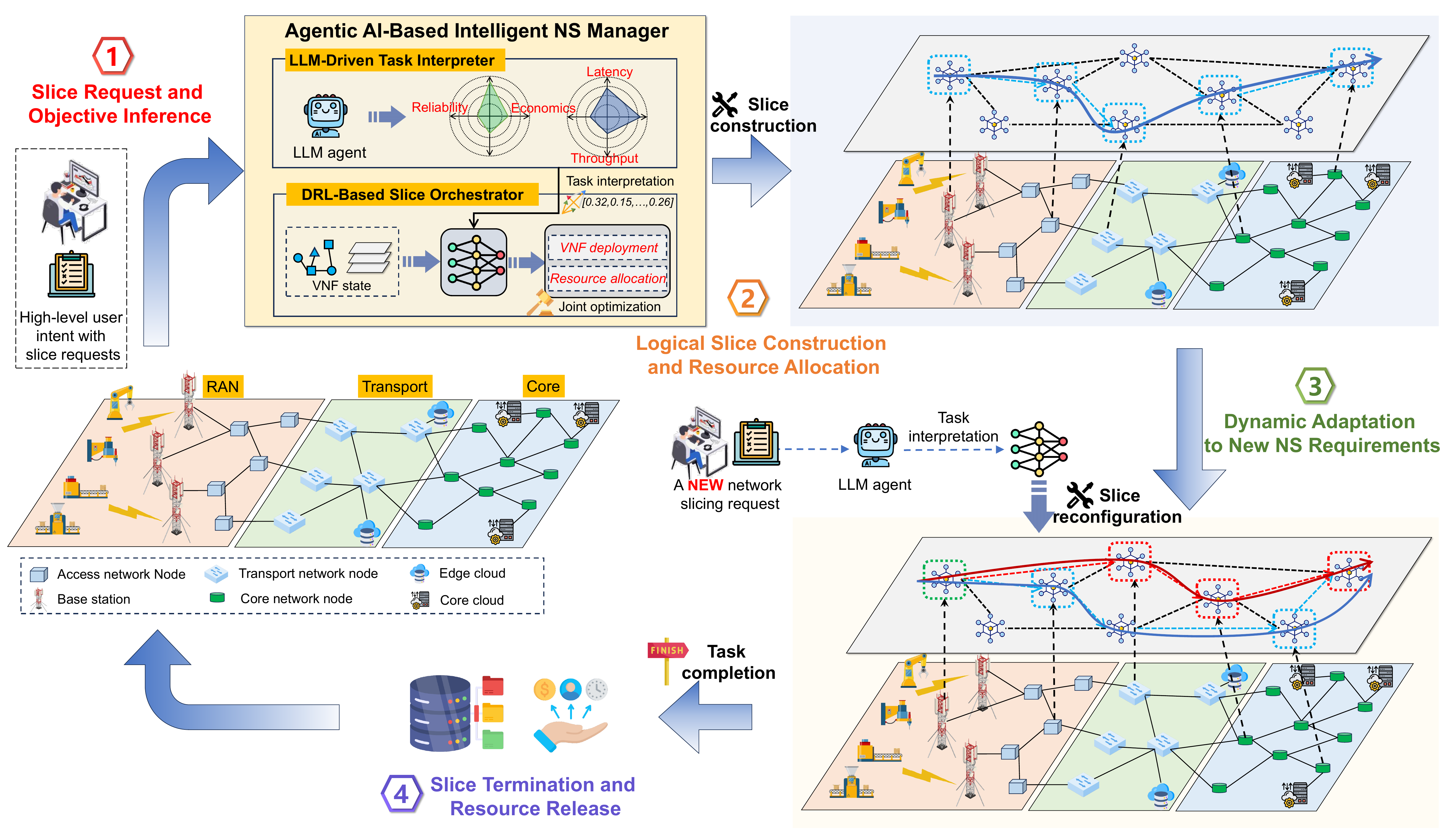}
    \caption{\small{The workflow of agentic AI-based intelligent network slice management. In step 1, slicing requests are issued, and the LLM-driven task interpreter generates task-specific performance weights. In step 2, the DRL-based slice orchestrator constructs a logical network slice and allocates physical resources accordingly. In step 3, the management system dynamically adapts to new slice requests through slice reconfiguration and VNF updates. In step 4, upon task completion, the slices are torn down and the allocated resources are released.}}
    \label{framework}
\end{figure*}

Since traditional optimization or heuristic methods are insufficient to meet the requirements of dynamic and complex environments, agentic AI has already shown strong potential in next-generation communication networks. Specifically, the authors in~\cite{10638533} demonstrated that LLM-enhanced multi-agent systems can overcome the limitations of conventional discriminative AI by introducing retrieval, collaborative planning, and reflection mechanisms. Such designs enable more adaptive and robust solutions for dynamic traffic variations, resource allocation, and semantic communication tasks. The authors in~\cite{xiao2025towards} proposed \textit{AgentNet}, a framework that transforms networking from a data-focused to a goal-oriented paradigm. By leveraging foundation-model-as-agents and embodied agents, \textit{AgentNet} supports decentralized autonomy, life-long learning, and secure collaboration, which are critical to achieving QoE-aware and resilient network management. In~\cite{11262045}, advanced architectures integrated with agentic AI were presented, highlighting constrained AI operations, autonomous cognitive agents, and neural radio protocol stacks. These innovations enable scalable, energy-efficient, and intent-driven network orchestration, reducing operational complexity and supporting sustainable service delivery.

Based on the above discussions, the distinct advantages of using agentic AI to address the challenges of dynamic IIoT NS management are summarized as below.
\begin{itemize}
    \item \textbf{Context-aware Intent Interpretation:} AI Agents equipped with context-awareness can interpret diverse service intents (e.g., power control and monitoring) and dynamically tailor network slice configurations to ensure the QoE satisfaction~\cite{10811956}.
    \item \textbf{QoE-driven Policy Mapping} Agentic AI can integrate network KPIs with application-level semantics, creating explainable models that translate latency, reliability, and cost metrics into actionable QoE-driven policies~\cite{10599123}.
    \item \textbf{Feedback-driven Proactive Slice Adaptation} By leveraging continuous feedback loops, agents can anticipate traffic bursts or workload shifts and proactively adjust slice allocations~\cite{jiang2025comprehensive}.
\end{itemize}

These advantages indicate that agentic AI offers a solid foundation for addressing the dynamic, heterogeneous, and QoE-sensitive nature of IIoT network slicing. Building on these insights, the next section presents our proposed LLM-empowered agentic AI approach for IIoT slicing management.

\section{Agentic AI-Based Network Slicing Management Framework Using LLMs}

In this section, we first propose the workflow of agentic AI-based intelligent network slicing management. Then, an LLM-empowered agentic AI approach for network slicing management is introduced to effectively support IIoT applications with heterogeneous QoE requirements.

\subsection{Workflow of Agentic AI-Based NS Management}

Network slicing enables the construction of logical networks over shared physical infrastructure to serve heterogeneous service demands. In the context of heterogeneous IIoT applications, the LLM and DRL agents are collaborated to provide intelligent, adaptive, and task-aware slice orchestration.  The detailed workflow of agentic AI-based adaption slice management is shown in Fig.~\ref{framework}.

\subsubsection{Slice Request and Objective Inference}The lifecycle begins when a high-level slicing request is issued by the NS administrator. The LLM agent acts as a task interpreter and analyzes natural-language task descriptions, as well as inferring the underlying performance requirements. Unlike fixed QoE templates, it dynamically assigns task-aware weights to a joint objective incorporating latency, reliability, and economics. For example, a mission-critical control loop may emphasize reliability, while a sensor fusion task may prioritize economics and update latency.

\subsubsection{Logical Slice Construction and Resource Allocation}Once task objectives are inferred, the DRL-based slice orchestrator constructs the logical slice and allocates physical resources. Guided by the QoE-aware reward design, the DRL agent determines the optimal deployment of VNFs, such as controllers, gateways, or processing nodes, while balancing performance and cost. Through agentic decision-making, the orchestrator performs joint VNF placement and resource allocation across RAN, transport, and core domains, ensuring QoE requirements while enhancing service-carrying capacity.

\subsubsection{Dynamic Adaptation to New NS Requirements}During runtime, application requirements may evolve due to varying workload intensity and environmental conditions. The agentic management system leverages real-time telemetry to monitor these changes. The LLM is then reinvoked to re-interpret updated requirements, while the DRL agent dynamically reconfigures the slice by updating VNFs.

\subsubsection{Slice Termination and Resource Release}Upon task completion, the agentic manager coordinates a graceful slice termination. All deployed VNFs are decommissioned or repurposed, and reserved resources are released to the shared infrastructure pool. This concludes the slice's lifecycle, allowing the infrastructure to be reused for future services.

In this workflow, the LLM is invoked only at the slice-episode or event level to infer high-level QoE objectives upon new or changed slice requests, while the DRL agent continuously performs per-decision-step orchestration within each episode based on the fixed LLM-derived preferences.

\subsection{LLM-Empowered Agentic AI Approach for NS Management}

\begin{figure*}
    \centering
    \includegraphics[width=0.95\linewidth]{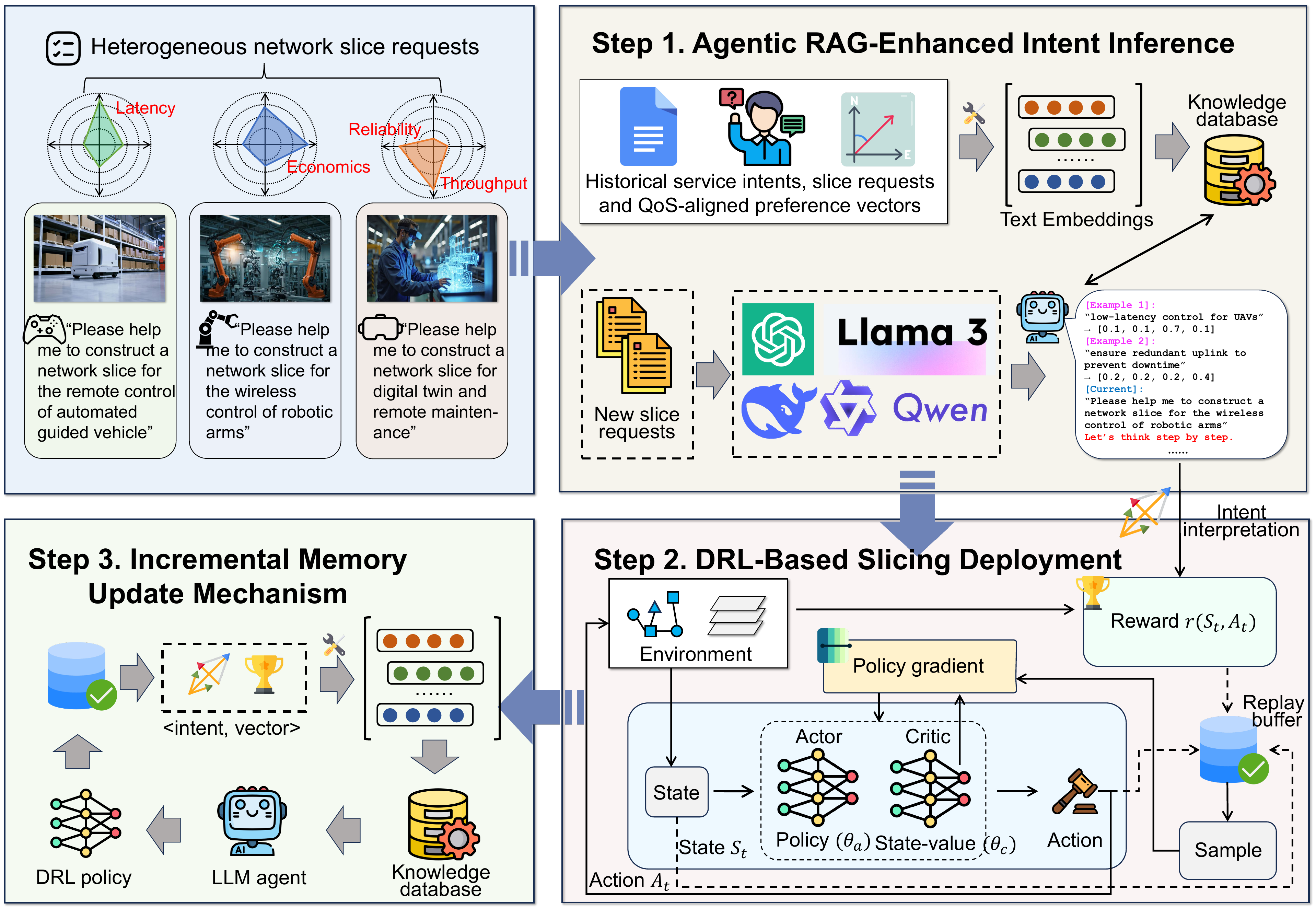}
    \caption{\small{The proposed LLM-driven network slicing management approach. In IIoT scenarios, heterogeneous applications initiate slice requests with distinct QoE preferences. In Step 1, the RAG-enhanced LLM module interprets user intents and outputs task-specific preference vectors reflecting latency, reliability, throughput, and control stability. In Step 2, the PPO-based DRL agent performs joint VNF placement and resource allocation based on network states and inferred preferences. In Step 3, the incremental memory update mechanism logs slicing outcomes and continually refines the semantic retrieval space, enabling self-evolving reasoning.}}
    \label{approach}
\end{figure*}

The proposed LLM-empowered agentic AI approach for network slicing management is illustrated in Fig.~\ref{approach}. First, the lightweight agentic RAG module interprets high-level slicing intents and translates them into task-aware preference vectors aligned with hybrid QoE performance objectives. Next, we design the DRL-based slice orchestrator to jointly determines VNF deployment strategy based on the inferred preferences and current network states. Finally, the incremental memory update mechanism continuously refines the intent-reasoning pipeline by logging feedback and enriching the knowledge database with newly observed performance outcomes. The details of each component are described as follows.

\subsubsection{\textbf{Step1: Agentic RAG-Enhanced Intent Inference}}
In IIoT slicing, accurate and low-latency interpretation of user-defined intents is vital for dynamic resource allocation under stringent QoE. To this end, we propose a lightweight RAG-based LLM agent that transforms natural-language requests into structured preference vectors aligned with hybrid QoE objectives. Specifically, the proposed RAG module centers around a lightweight semantic retrieval engine co-located with the edge LLM. In practice, the edge LLM is deployed at the edge cloud rather than on IIoT end devices, enabling low-latency intent inference while remaining compatible with realistic computational constraints. Historical service intents, user requests, and corresponding QoE-aligned preference vectors are embedded and indexed into a vector database using a compact transformer encoder DistilBERT. Each entry in the database is represented as a pair $<\mathrm{intent\_text}, \mathrm{preference\_vector}>$, where the vector encodes the relative importance across core IIoT slicing dimensions such as latency, reliability, throughput, and economics. Upon receiving a new user prompt $\mathbf{p}$, the RAG module computes its semantic embedding and performs a top-k nearest neighbor search within the vector space. The retrieved examples $\left\{\left(\mathbf{p}_{i}, \mathbf{s}_{i}\right)\right\}_{i=1}^{k}$ serve as anchors that ground the model's reasoning in previously observed behaviors. This enables edge-hosted LLMs (e.g., GPT-4o, LLaMA 3, and DeepSeek) to perform plug-and-play intent inference without retraining. The retrieved context is formatted into a few-shot prompt combining past pairs and the current query. The edge LLM, with as few as 1–2 billion parameters, processes the prompt and outputs a preference vector $\hat{\mathbf{s}}$ aligned with QoE attributes, which serves as an intent embedding passed downstream to guide DRL decision-making.

\subsubsection{\textbf{Step 2: DRL-Based Slicing Deployment}}
The deployment of network slices is managed by a DRL agent, specifically VNF placement and resource allocation. At each step, the agent observes a structured state capturing both network status and user expectations, including real-time computing power, latency on candidate VNF nodes, and slice topology. Crucially, the state incorporates a QoE preference vector from the upstream LLM agent, quantifying user emphasis on latency, reliability, and economics. As for action space, the agent selects an appropriate set of VNF nodes from a pool of available VNF resources. A multi-objective reward function, dynamically weighted by the LLM-derived preferences, aligns the agent's policy updates with heterogeneous service requirements. Instead of relying on a fixed optimization goal, the importance of different performance dimensions is weighted according to the semantic vector provided by the LLM.

\subsubsection{\textbf{Step 3: Incremental Memory Update Mechanism}} To ensure long-term adaptability under dynamic network and application conditions, the proposed network slicing management approach incorporates an incremental memory update mechanism that supports continual refinement of inference quality without requiring model retraining. This mechanism consists of a knowledge database, a retrieval store, and a memory bank, where the retrieval store enables semantic indexing and fast retrieval of historical intents, while the memory bank incrementally accumulates performance feedback and task outcomes for lifelong learning. This mechanism is built atop the RAG backbone and operates through structured logging, intent-performance association, and index updating. Specifically, after each slicing decision, the observed performance feedback and QoE metrics are logged and transformed into new $<\text{intent}, \text{vector}>$ entries. These entries are then normalized, time-stamped, and encoded into compact vector representations using a shared embedding model consistent with the retrieval store's schema. Once encoded, they are indexed back into the RAG memory bank via a vector similarity-based insertion policy, enabling lifelong learning without retraining. To avoid abrupt forgetting while maintaining memory compactness, the memory bank adopts a time-stamped soft aging strategy, where newly inserted entries are prioritized during retrieval, while outdated or less relevant records are implicitly down-weighted rather than explicitly deleted. In addition, a similarity-based redundancy control mechanism is employed, such that highly redundant experiences exceeding a predefined similarity threshold are merged or summarized instead of being stored as independent entries. During future slicing decisions, the LLM component performs semantic retrieval over this memory bank. Because the store is continuously updated, the LLM's context window is enriched with more recent and situationally relevant deployment outcomes. This forms a closed semantic-control feedback loop, wherein past decisions inform future reasoning paths. In practical deployments, the initial entries of the knowledge database are bootstrapped from historical network slicing logs, where previously issued slice requests, their associated service descriptions, and operator-defined SLA/QoE configurations naturally form structured ⟨intent, vector⟩ pairs.
% For example, if the system detects suboptimal QoE for intents emphasizing reliability performance, the updated examples implicitly shift the retrieved context toward latency-sensitive responses in future queries, thereby optimizing inference quality in a self-correcting fashion.
Note that the memory update and retrieval process is driven by semantic similarity and vector similarity-based insertion, rather than intent frequency, thereby mitigating potential bias toward frequently occurring intents.

\section{Case Study: VNF Deployment for Heterogeneous Network Slicing}

\subsection{System Model}

We consider a case study involving heterogeneous network slice requests, where multiple small-cell base stations forward IIoT traffic to a central unit (CU) responsible for managing and orchestrating VNFs across network slices. Each slice, requested sequentially by users, corresponds to a distinct IIoT service class with specific QoE requirements in terms of latency, reliability, and economics. The CU operates a pool of servers with limited CPU and memory resources, and relies on lightweight containerization to deploy VNF instances for different slices. When local capacity is insufficient, the CU can offload VNFs to a central cloud with virtually unlimited resources, but at the expense of higher delay and financial cost. Thus, the CU dynamically orchestrates VNFs for each slice through vertical scaling of existing resources, horizontal scaling via new container instantiation, or cloud offloading to maintain QoE of slices under fluctuating IIoT traffic demands.

% This architecture leads naturally to a multi-objective optimization problem that captures the trade-offs between latency, cost, and reliability across slices. Latency costs include container deployment delays and delays of VNF nodes, as well as cloud offloading overhead; financial costs arise from server usage and cloud service charges; and the reliability metric is measured by the number of VNF nodes shared with other slices within an end-to-end network slice, i.e., the more shared nodes there are, the lower the reliability. These objectives are inherently conflicting: for instance, minimizing latency for URLLC slices may require resource over-provisioning, raising operational cost, while aggressively reducing cost could result in degraded reliability for mission-critical industrial control. To balance these objectives, the system defines a weighted cost function integrating latency, cost, and reliability, enabling operators to adjust priorities across different IIoT slices. In this formulation, the latency component, weighted by $\alpha$, quantifies the delays due to container initialization, VNF nodes, and cloud offloading; the cost component, weighted by $\beta$, reflects the resource consumption at local servers and deployment of VNF nodes, as well as the charges incurred when using cloud resources; and the reliability penalty component reflects the number of VNF nodes shared by all network slices in the system.
This architecture naturally leads to a multi-objective optimization problem capturing trade-offs among latency, cost, and reliability across slices. Latency costs include container deployment, VNF node delays, and cloud offloading; financial costs arise from server usage and cloud service charges; reliability is measured by the number of VNF nodes shared across slices, specifically, the reliability cost is proportional to the number of shared nodes. These objectives conflict: minimizing latency for URLLC may require resource over-provisioning, raising cost, while aggressive cost reduction may degrade reliability for mission-critical control. To balance these factors, the system defines a weighted cost function that integrates latency, economic cost, and reliability, where these metrics are of the same order of magnitude, thereby ensuring that no single metric is favored due to scale differences. Here, the weighted latency component quantifies delays from container initialization, VNF nodes, and offloading; the weighted cost component reflects local server use, VNF deployment, and cloud charges; and the weighted reliability component reflects VNF node sharing across slices.

\subsection{Experimental Configuration}

% To ensure reproducibility of our results, we carefully specify the experimental setup and parameter configurations. The system consists of 4 $\sim$ 20 slice requests with different QoE, a CU equipped with a local server and a cloud server with powerful computational capabilities. Local server provides a maximum of 40 CPU units and 30 memory units, and cloud server provides a unlimited CPU and memory units. Each network slice requires different CPU and memory resources; for instance, one slice demands 2 to 5 CPU units and 2 to 4 memory units, while another requires 3 to 6 units of both CPU and memory. The container boot-up delay is 30 milliseconds. The costs of local and cloud server are respectively set to 30 and 10 units. The local server is connected to the cloud server through the deployment of VNFs, with each VNF incurring a deployment delay of 10 to 15 milliseconds and an associated cost of 2 to 4 units.
To ensure reproducibility, we specify the experimental setup and parameters. The system handles 4$\sim$20 slice requests with diverse QoE over a period of time, and the system consists of a CU with a local server, a cloud server with powerful computation, and a set of VNF nodes which is used for slice construction. The QoE of each network slice is quantified as a weighted aggregation of latency, reliability, and economics, reflecting task-specific performance priorities. Note that high-level slice requests with different QoE requirements occur sequentially during this period. The local server provides up to 40 CPU and 30 memory units, while the cloud server offers unlimited resources. Each slice requires varying CPU and memory; for example, one demands 2$\sim$5 CPU and 2$\sim$4 memory units, while another requires 3$\sim$6 units of both. The container boot-up delay is 30 ms. The local and cloud server costs are 30 and 10 units, respectively. The local server is connected to the cloud server through the deployment of VNFs, with each VNF incurring a deployment delay of 10$\sim$15 milliseconds and an associated cost of 2$\sim$4 units. The end-to-end latency is constrained to be no greater than 30 ms for high-priority slices, 100 ms for medium-priority slices, and 150 ms for best-effort slices. Besides, high-reliability slices allow at most 1 slice per VNF node, medium-reliability slices allow at most 2 slices per VNF node, and best-effort slices allow up to 4. The economic constraint is set to 25 units for cost-sensitive slices and 40 units for general-purpose slices. A slice is considered successfully served only if all these constraints are simultaneously satisfied, which defines the slice availability ratio used in the experimental results.

Based on the above LLM-driven framework, we propose a QoE-aware Proximal Policy Optimization (QAPPO) method to address VNF deployment under heterogeneous slicing. The learning rate for both actor and critic networks are set to $1\times10^{-4}$. The discount factor is set to $0.99$. The clipping parameter is set to $0.1$, and the batch size is set to 1024. For fair comparison, the experiments also include the following baselines: a pure PPO approach without LLM-based QoE awareness, a Local-First strategy, and a Cloud-Only strategy.
% All experiments are conducted on an Ubuntu server equipped with a 2.10 GHz Intel Xeon Gold 5218R CPU (40 cores, 503 GB RAM) and an NVIDIA A100 80 GB GPU. 
For LLMs, we integrate GPT-4o via the OpenAI API as a pluggable module for intent inference and QoE weight assignment. Besides, the knowledge database and context memory are built using LangChain to enable RAG function. 
% The retrieval memory is constructed using historical service intents and their corresponding QoS preference vectors, embedded via a compact transformer encoder DistilBERT and indexed in a vector database.

\subsection{Experimental Results}

\begin{figure}[!t]
\centering
\includegraphics[width=0.85\linewidth]{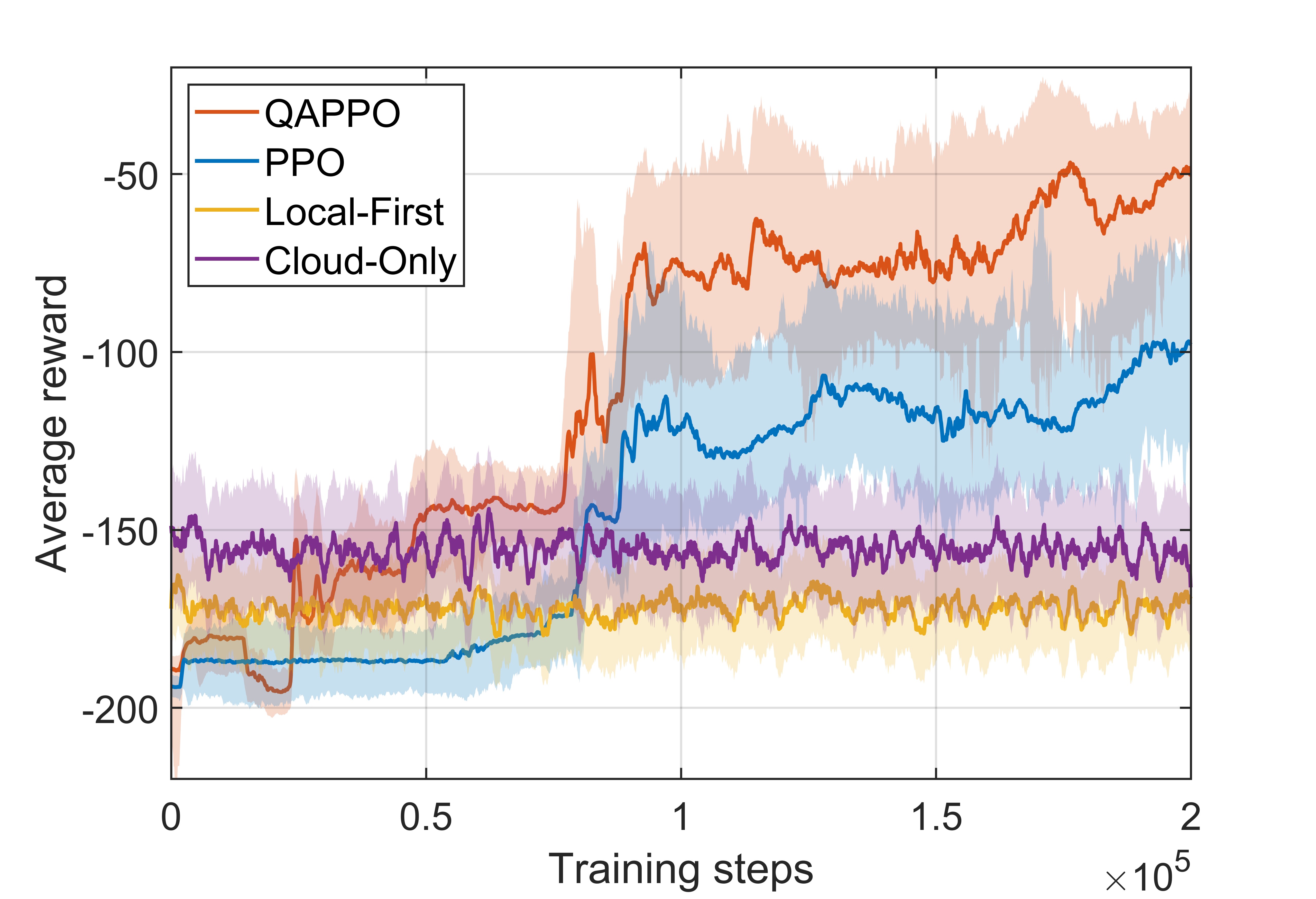}
\caption{The reward curves of different methods.}
\label{training_curve}
\end{figure}

Fig.~\ref{training_curve} presents the training reward curves of different methods. It is evident that the proposed QAPPO achieves the highest long-term reward, significantly outperforming the conventional PPO, Local-First, and Cloud-Only baselines. And QAPPO exhibits a sharp performance gain after 75k training steps. Specifically, the average reward of QAPPO consistently outperforms PPO, indicating the effectiveness of the LLM-driven intent interpretation and adaptive weight assignment in guiding the DRL agent. By contrast, PPO without LLM support suffers from lower reward due to its reliance on static and manually assigned weights. The heuristic strategies, including Local-First and Cloud-Only, remain nearly flat with significantly lower rewards, as they lack the ability to dynamically balance latency, cost, and reliability trade-offs, leading to an inability to accommodate heterogeneous slice requirements.

\begin{figure*}[t]
\centering
    \begin{subfigure}[b]{0.26\textwidth}
            \includegraphics[width=\textwidth]{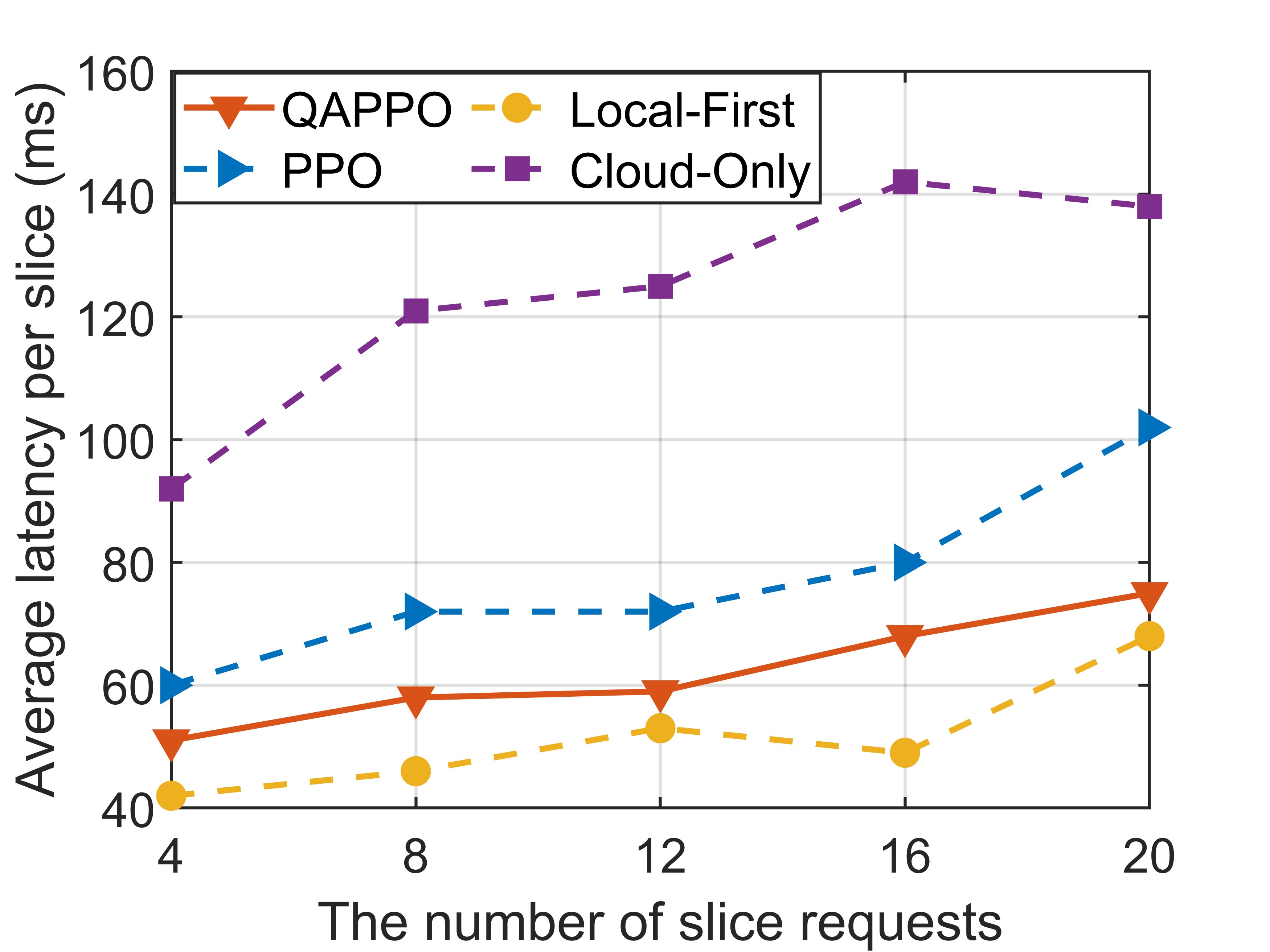}
            \subcaption{}
            \label{latency_vs_user}
    \end{subfigure}
    %\hfil
    \hspace{-6mm}
    \begin{subfigure}[b]{0.26\textwidth}
            \includegraphics[width=\textwidth]{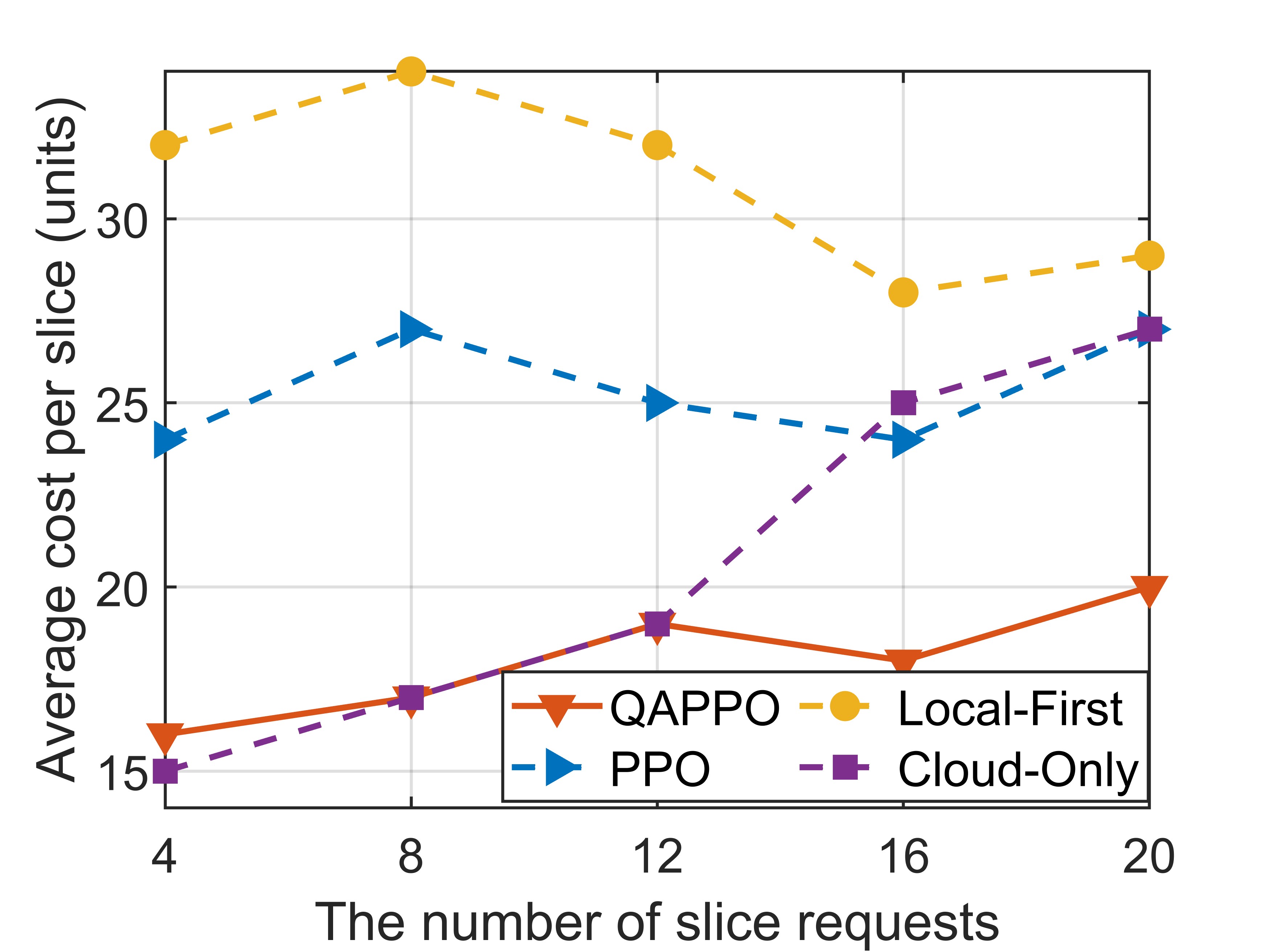}
            \subcaption{}
            \label{cost_vs_user}
    \end{subfigure}
    %\hfil
    \hspace{-6mm}
    \begin{subfigure}[b]{0.26\textwidth}
            \includegraphics[width=\textwidth]{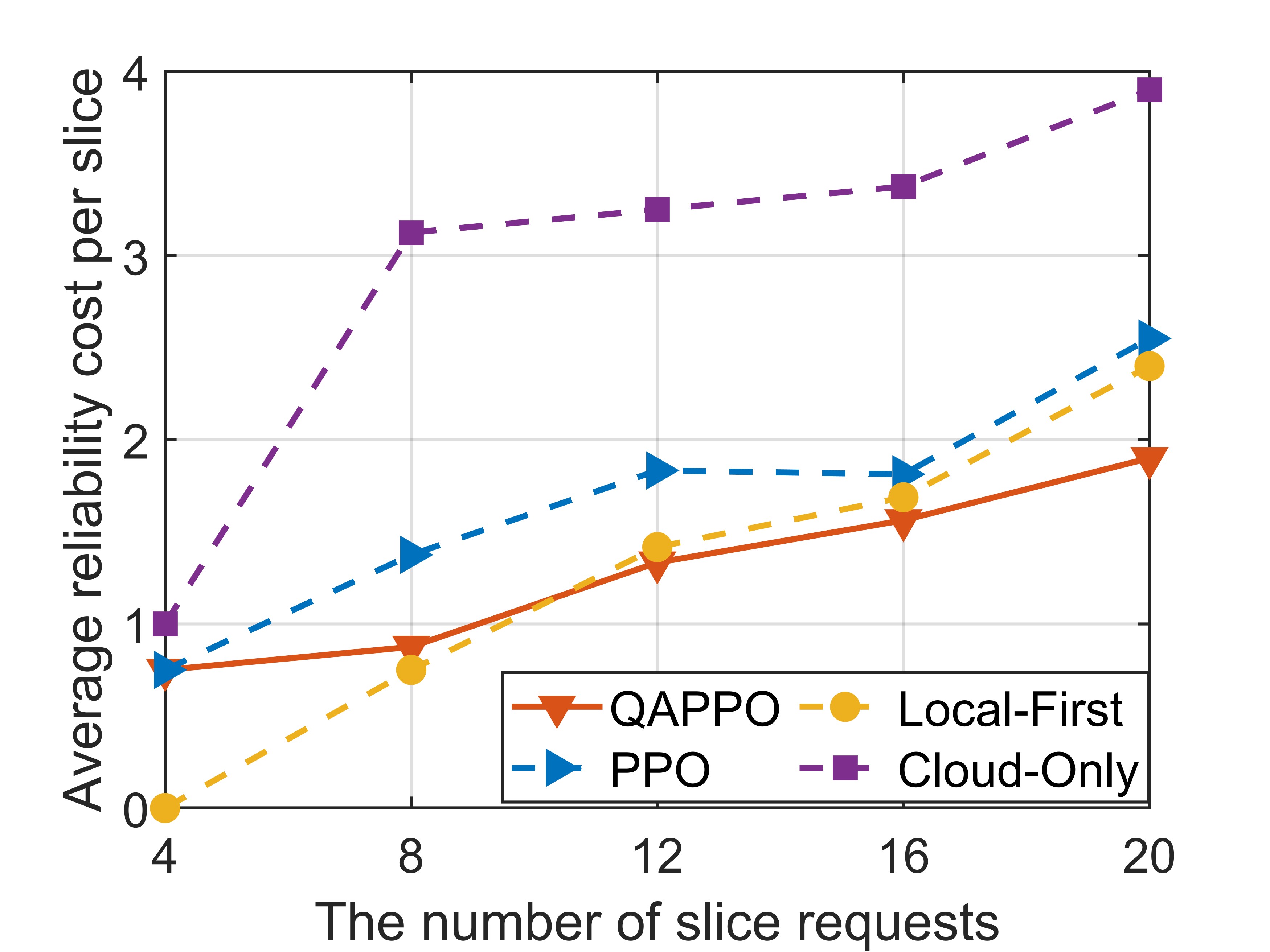}
            \subcaption{}
            \label{reliability_cost_vs_user}
    \end{subfigure}
    %\hfil
    \hspace{-6mm}
    \begin{subfigure}[b]{0.26\textwidth}
            \includegraphics[width=\textwidth]{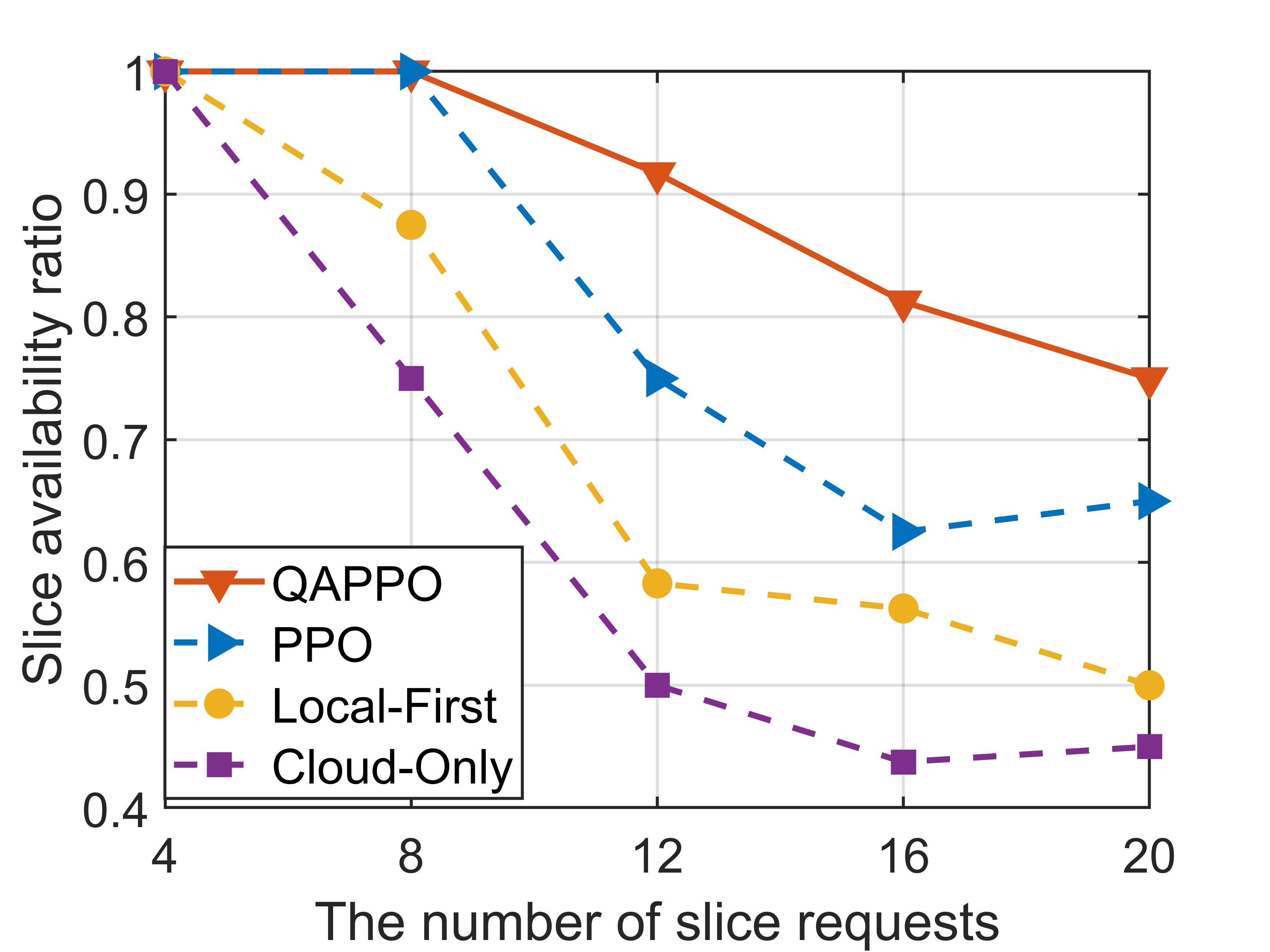}
            \subcaption{}
            \label{slice_availability_ratio}
    \end{subfigure}
    
\caption{\small{Experimental results under varying numbers of network slice requests over a period of time: (a) average latency per slice versus the number of slice requests; (b) average cost per slice versus the number of slice requests; (c) average reliability cost per slice versus the number of slice requests; (d) slice availability ratio versus the number of slice requests. Note that the slice availability ratio is calculated as the proportion of slice requests that satisfy the QoE constraints over the total number of slice requests.}}
\label{experiment_results}
\end{figure*}

To further examine system-level QoE, Fig.~\ref{experiment_results}(a)-(c) compares the performance of all schemes under varying numbers of slice requests.
We observe that the proposed QAPPO method achieves the lowest average cost and reliability cost, as well as the second-lowest latency. This demonstrates that by continuously perceiving slice requests and assigning adaptive weights, the LLM agent enables QAPPO to intelligently orchestrate VNF nodes between local and cloud servers while minimizing VNF sharing across slices. 
% The Local-First baseline achieves the lowest average latency per slice across varying numbers of slice requests. This is because data flows are prioritized for processing on the local server without traversing VNF nodes. In addition, the Local-First method yields the lowest reliability cost when the number of slice requests is small; however, as the request volume increases, its reliability cost surpasses that of QAPPO. This is because the limited resources of the local server force Local-First to offload tasks to the cloud, where its scheduling efficiency is inferior to that of QAPPO. The Cloud-Only baseline achieves the lowest average cost when the number of slice requests is low, but the cost increases steadily as the number of requests grows. This is because, with more slice requests, the Cloud-Only approach must deploy additional nodes to ensure that the number of shared VNF nodes does not exceed the constraints, thereby meeting the reliability requirements of each slice.
The Local-First baseline achieves the lowest average latency per slice across varying slice request volumes, as data flows are prioritized for processing on the local server without traversing VNF nodes. It also yields the lowest reliability cost when slices requests are few; however, as demand rises, its reliability cost surpasses QAPPO due to limited local resources and less efficient cloud offloading. The Cloud-Only baseline achieves the lowest average cost with few slice requests, but cost increases steadily as slice requests grow, since additional nodes must be deployed to meet reliability constraints of each slice by limiting VNF node sharing.
Furthermore, due to the lack of dynamic and task-specific weight assignment, the PPO baseline lags behind QAPPO in terms of latency, average cost, and reliability performance.

Finally, Fig.~\ref{experiment_results}(d) reports the slice availability ratio, defined as the proportion of requests that satisfy QoE constraints. QAPPO achieves the highest availability, sustaining over $75\%$ even with $20$ slice requests, whereas PPO gradually degrades to about $65\%$. Local-First and Cloud-Only fall below $50\%$, indicating frequent violations of QoE guarantees under high demand. These results demonstrate that our proposed QAPPO effectively leverages LLM-driven QoE awareness to achieve superior trade-offs among latency, cost, reliability, thereby ensuring robust performance for heterogeneous IIoT services, especially under conditions of slice request overload and constrained network service resources. Notably, with $16$ slice requests, QAPPO achieves up to a $19\%$ improvement in availability over baselines, demonstrating strong scalability of the agentic AI approach in dynamic management.

\section{Future Directions}

% \textbf{Self-Evolving Memory and Lifelong Adaptation:} Agentic AI for IIoT slicing must continuously evolve with dynamic industrial contexts. Future work should move beyond static RAG modules toward self-evolving memory architectures where agents not only store but also abstract, generalize, and compress long-term experiences. Mechanisms such as neural-symbolic memory fusion, meta-learning-driven memory consolidation, and continual self-refinement will allow slicing agents to accumulate domain expertise and autonomously adapt to unseen QoE requirements.
\textbf{Self-Evolving Memory and Lifelong Adaptation:} Agentic AI for IIoT slicing should transition from static RAG designs to self-evolving memory architectures that support abstraction, generalization, and compression of long-term experiences. By integrating neural–symbolic memory, meta-learning–based consolidation, and continual self-refinement, slicing agents can progressively accumulate domain knowledge and autonomously adapt to unseen and evolving QoE requirements.

% \textbf{Multi-Agent Collaboration and Negotiation Mechanisms:}
% Industrial networks involve multiple stakeholders and heterogeneous slices, often with conflicting QoE objectives. Future directions should explore decentralized agentic ecosystems where multiple LLM-driven agents collaborate, negotiate, and reach consensus on resource allocation. Incorporating game-theoretic reasoning, contract-based negotiation, and decentralized trust mechanisms could significantly improve fairness, stability, and scalability of slice orchestration in multi-tenant IIoT environments.
\textbf{Multi-Agent Collaboration and Negotiation Mechanisms:} Multi-tenant IIoT networks feature heterogeneous slices and competing QoE objectives. Decentralized multi-agent frameworks, where LLM-driven agents collaborate and negotiate resource allocation, represent a promising direction. Game-theoretic reasoning, contract-based negotiation, and decentralized trust mechanisms can enhance fairness, stability, and scalability in slice orchestration.

% \textbf{Embodied Agentic AI for Cross-Layer Orchestration:}
% Most existing designs position agentic AI at the control or management plane. A promising direction is to endow agents with embodied capabilities across the physical, network, and application layers. For example, embodied agents could sense radio conditions, reason about cross-layer trade-offs, and directly reconfigure slice topologies in real time. Techniques such as neural radio protocol stacks, intent-driven cross-layer reasoning, and embodied decision feedback loops could enable more holistic and adaptive QoE optimization.
\textbf{Embodied Agentic AI for Cross-Layer Orchestration:} Beyond management-plane intelligence, embodied agentic AI can operate across physical, network, and application layers. By sensing radio conditions, reasoning over cross-layer trade-offs, and directly reconfiguring slice topologies, such agents enable real-time, holistic QoE optimization. Neural radio stacks, intent-driven reasoning, and closed-loop decision feedback are key enablers of this paradigm.

\section{CONCLUSION}

We have presented a forward-looking perspective on LLM-empowered agentic AI for QoE-aware network slicing management in Industrial IoT. We introduced the IIoT slicing architecture and relevant QoE metrics, and analyzed the challenges of dynamic slice management along with the motivations and advantages of employing agentic AI. Building on this, we proposed an LLM-empowered agentic AI approach for network slicing management, integrating retrieval-augmented intent inference, DRL-based orchestration, and incremental memory, and showed via a VNF case study its superiority in balancing performance and cost under dynamic workloads. Looking ahead, LLM-empowered agentic AI is expected to be a cornerstone for IIoT and 6G, enabling autonomous, explainable, and adaptive network slicing for digital twins, cyber-physical systems, and intelligent automation.

%\printbibliography % 对应 package:biblatex

% \bibliographystyle{ieeetr} % 会导致number 和 month 只显示 month
\bibliographystyle{IEEEtran}
\bibliography{references}

% \section*{BIOGRAPHIES}  %不带序号
% \textbf{Xudong Wang} is a PhD candidate at Beijing University of Posts and Telecommunications, China. \textbf{Jiacheng Wang} is a research fellow at Nanyang Technological University, Singapore. \textbf{Lei Feng} is an associate professor with Beijing University of Posts and Telecommunications, China. \textbf{Dusit Niyato} is a chair professor with Nanyang Technological University, Singapore. \textbf{Ruichen Zhang} is a research fellow at Nanyang Technological University, Singapore. \textbf{Jiawen Kang} is a professor with Guangdong University of Technology, China. \textbf{Zehui Xiong} is an assistant professor with Singapore University of Technology and Design, Singapore. \textbf{Hongyang Du} is an assistant professor with The University of Hong Kong, Hong Kong. \textbf{Shiwen Mao} is a professor with Auburn University, Auburn, USA.

%\vfill

\end{document}